\documentclass[aps,prd,preprint,superscriptaddress,amsmath,amssymb,showpacs]{revtex4-1}
\usepackage{dcolumn}
\usepackage{graphicx}
\usepackage{float}
\usepackage{physics}
\usepackage[colorlinks=true,allcolors=blue]{hyperref}

\begin{document}

\title{Moving heavy quarkonium entropy, effective string tension, and the QCD phase diagram}

\author{Xun Chen}
\affiliation{College of Science, China Three Gorges University, Yichang 443002, China}

\author{Sheng-Qin Feng}
\email{fengsq@ctgu.edu.cn}
\affiliation{College of Science, China Three Gorges University, Yichang 443002, China}
\affiliation{Key Laboratory of Quark and Lepton Physics (MOE) and Institute of Particle Physics,\\
Central China Normal University, Wuhan 430079, China}

\author{Ya-Fei Shi}
\affiliation{College of Science, China Three Gorges University, Yichang 443002, China}

\author{Yang Zhong}
\affiliation{College of Science, China Three Gorges University, Yichang 443002, China}


\date{\today}

\begin{abstract}
The entropy and effective string tension of the moving heavy quark-antiquark pair in the strongly coupled plasmas are calculated by using a deformed an anti-de Sitter/Reissner-Nordstr$\ddot{o}$m black hole metric. A sharp peak of the heavy-quarkonium entropy around the deconfinement transition can be realized in our model, which is consistent with the lattice QCD result. The effective string tension of the heavy quark-antiquark pair is related to the deconfinement phase transition. Thus, we investigate the deconfinement phase transition by analyzing the characteristics of the effective string tension with different temperatures, chemical potentials, and rapidities. It is found that  the results of phase diagram calculated through effective string tension are in agreement with results calculated through a Polyakov loop. We argue that a moving system will reach the phase transition point at a lower temperature and chemical potential than a stationary system. It means that the lifetime of the moving QGP becomes longer than the static one.
\end{abstract}



\maketitle

\section{Introduction}\label{sec:01_intro}
The study of a strongly coupled Yang-Mills theory, such as quantum chromodynamics(QCD), is still a challenge in spite of the methods developed so far. The string description of realistic QCD has not been successfully formulated yet. Many ``top-down'' methods are invested in searching for such a realistic description by deriving holographic QCD from the string theory\cite{Huang:2007fv,Kruczenski:2003uq,Sakai:2004cn,Sakai:2005yt}. On the other hand, the ``bottom-up'' approach has been used to examine possible holographic QCD models from experimental data and lattice results. A black hole in  five-dimensional space is induced to depict the boundary theory at a finite temperature \cite{Witten:1998zw, Brandhuber:1998bs, Rey:1998bq} and also to discuss more general backgrounds\cite{Kinar:1998vq}. In the bottom-up approach, the most economic way is to use a deformed anti-de Sitter ($\mathrm{AdS}_5$) metric\cite{He:2010ye,Pirner:2009gr,Shock:2006gt,Zeng:2008sx,Brodsky:2010ur,Ghoroku:2003ex,Gursoy:2007er}, which can describe the known experimental data and lattice results of QCD, e.g., hadron spectra and the heavy-quark potential.

Since QCD is not a conformal theory, a mechanism for breaking conformal invariance must be included in the dual model. The deformed $\mathrm{AdS}_5$ metric corresponding to super Yang-Mills(SYM)\cite{Maldacena:1998im} multiplied by an overall warp factor is introduced in the soft-wall model. The soft wall model at a finite temperature has been introduced in Refs.~\cite{Karch:2006pv,Andreev:2006eh}, which was motivated by the zero temperature soft wall model\cite{Kajantie:2006hv} that achieved considerable success in describing various aspects of hadron physics.

QCD is an asymptotically free theory; thus, its high temperature and high density phases are dominated by quarks and gluons as degrees of freedom. These phases play an important role in relativistic heavy ion collisions. Matsui and Satz\cite{Matsui:1986dk} indicated that the binding interaction of the heavy $Q\bar{Q}$ pair is screened by the quark gluon plasma (QGP) medium, leading to the melting of the heavy quarkonium. However, the $Q\bar{Q}$ pair is unlikely generated at rest in the QGP, and the effects of its motion through the plasma must be considered when taking into account the effects of the medium in the $Q\bar{Q}$ interaction. It is found that the heavy quarkonia have finite probability to survive even at infinitely high temperature. The study of moving heavy quarkonium in QGP from the effective field theory has been done in Refs.~\cite{Song:2007gm,Escobedo:2013tca}.

The effect of a finite quark density in QCD, on the other hand, is induced by adding the term $J_D = \mu \psi^\dagger(x) \psi(x)$ to the Lagrangian in the generating functional. Therefore, the chemical potential $\mu$ is characterized by the source of the quark density operator. The source of a QCD operator in the generating functional is related to the boundary value of a dual field in the bulk according to the AdS/CFT correspondence. Therefore, the chemical potential can be treated as the boundary value of the time component of a $\mathrm{U}(1)$ gauge field $A_M$ dual to the vector quark current. The solution is known as the AdS/Reissner-Nordstr$\ddot{o}$m(AdS/RN) black hole metric and describes a charged black hole interacting with the electromagnetic field\cite{Jo:2009xr,Chamblin:1999hg,Lee:2009bya}. To describe the finite temperature and density in the boundary theory, the dual space geometry of anti-de Sitter with a charged black hole is introduced in Ref.\cite{Colangelo:2010pe} with a background warp factor.

References.~\cite{Andreev:2006eh,Andreev:2006nw} used gauge/string duality to investigate the free energy of a heavy $Q\bar{Q}$  pair in strongly interacting matter and introduced the notion of an effective string tension at a finite temperature. As we know, the free energy includes the attractive potential and repulsive potential. The effective string tension comes from the calculation of the free energy. Reference\cite{Andreev:2006nw} has used the effective string tension to discuss the phase transition problem.

The phase diagrams in the $\mu$--$T$ plane and entropy are investigated by considering a moving heavy $Q\bar{Q}$  pair in our article. The paper is organized as follows. In Sec.~\ref{sec:02_setup}, we introduce a deformed AdS/RN black hole metric setup. In Sec.~\ref{sec:03}, we calculate the effective string tension and phase diagram in the $\mu$--$T$ plane. The entropy of a moving quarkonium is computed in Sec.~\ref{sec:04}. We give a short discussion and conclusion in Sec.~\ref{sec:05}.

\section{The Setup}\label{sec:02_setup}

The solution of the equation of motion of a 5D gravity action with a negative cosmological constant interacting with an electromagnetic field is known as an AdS/RN black hole metric\cite{Colangelo:2010pe},
and we consider the deformed AdS/RN black hole metric,
\begin{gather}
\dd{s}^2 = \frac{R^2 h(z)}{z^2}(-f(z) \dd{t}^2 + \dd{x}^2 + \frac{\dd{z}^2}{f(z)}) \\
f(z) = 1 - \qty(\frac{1}{z_h^4} + q^2 z_h^2) z^4 + q^2 z^6,
\end{gather}
where $h(z) = \exp(cz^2/2)$ is called as the warp factor, which determines the characteristics of the soft wall model, and the deformation parameter $c$ determines the deviation from conformality\cite{Andreev:2006ct,Andreev:2010bv}.
In Eq.~(2), $q$ is the black hole charge, $z_h$ is the position of black hole horizon, $R$ is the radius of curvature (set to 1), $x$ is the spatial directions of the space-time, and $z$ is the fifth holographic coordinate. The Hawking temperature of the black hole is defined as
\begin{equation}
T = \frac{1}{4\pi} \qty|\frac{\dd{f}}{\dd{z}}|_{z = z_h} = \frac{1}{\pi z_h} \qty(1 - \frac{1}{2} Q^2),
\end{equation}
where $Q = qz_h^3$ and $0 \leq Q \leq \sqrt{2}$. The relationship between the chemical potential $\mu$ and $q$ is given as,
\begin{equation}
\mu = \kappa \frac{Q}{z_h},
\end{equation}
where $\kappa$ is a dimensionless parameter, and we fix the parameter $\kappa$ to one in this paper.

\begin{figure}
    \centering
    \includegraphics[width=8.5cm]{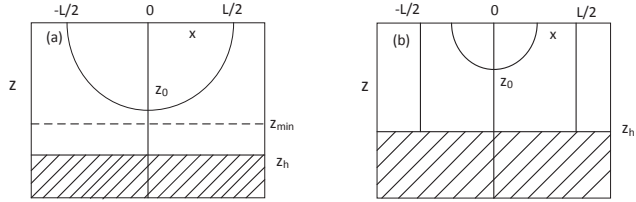}
    \caption{\label{fig1}(a) shows the feature of a U-shape open string connects the $Q\bar{Q}$  in the confinement phase, and (b) shows two straight strings reach the position of horizon at large separating distances in the deconfinement phase.}
\end{figure}

\begin{figure}
    \centering
    \includegraphics[width=8.5cm]{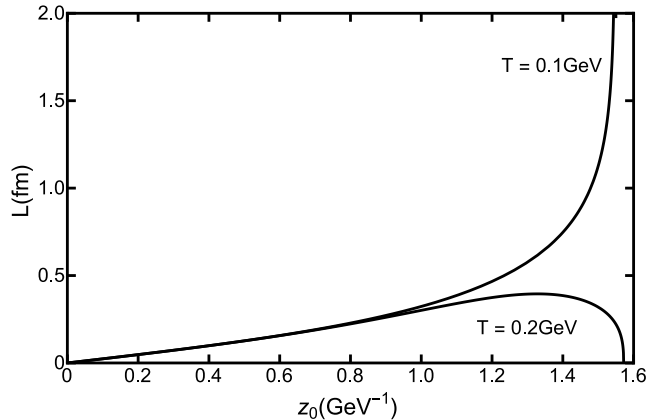}
    \caption{\label{fig2}The dependences of interquark distance $L$ of $Q\bar{Q}$  pair on $z_0$ at the low temperature $T = 0.1\,\mathrm{GeV}$ and at the high temperature $T = 0.2\,\mathrm{GeV}$ for a given chemical potential $\mu = 0.1\,\mathrm{GeV}$.}
\end{figure}

A fundamental string connects the $Q\bar{Q}$  as shown in Fig.~\ref{fig1}(a), and a U-shape open string connects the $Q\bar{Q}$  in the confinement phase. There is a dynamic wall $z_m$ at a low temperature, and the U-shape string can not exceed the dynamic wall with increasing the separating distances of the $Q\bar{Q}$  pair as shown in the situation of $T = 0.1\,\mathrm{GeV}$ in Fig.~\ref{fig2}. When the temperature increases to a certain value (critical temperature) and the value of horizon distance $z_h$ becomes small, the system will undergo a confinement-deconfinement phase transition. Figure~\ref{fig1}(b) shows the deconfinement phase, in which the U-shape string can reach the position of the horizon and become two straight strings at large separating distances. In the situation, there is a large black hole, and the dynamic wall disappears. The critical temperature $T_c$ can be interpreted as a confinement-deconfinement phase transition temperature. In the deconfinement phase, if the separating distances $L$ of $Q\bar{Q}$  pair is small, the U-shape string still exists.

 When $z_0$ increases, the separating distance $L$ of the $Q\bar{Q}$  pair reaches the maximum. When
 the separating distance of $Q\bar{Q}$  pair is larger than the maximum, U-shape strings become unstable and finally develop two straight open strings connecting the boundary and horizon as shown in the situation of $T = 0.2\,\mathrm{GeV}$ in Fig.~\ref{fig2}.

The study of a moving quarkonium of an AdS/CFT correspondence has been carried out in Refs.~\cite{Fadafan:2015ynz,Fadafan:2015kma}, but not in the deformed metric. We will study the moving quarkonium in the deformed AdS/RN metric. It is well known that the quarkonium is moving with relativistic velocities through the QGP in relativistic heavy ion collisions. We consider the general background (Minkowski metric),

\begin{equation}
    \dd{s}^2 = -G_{tt} \dd{t}^2 + G_{xx}\dd{x_i}^2 + G_{zz}\dd{z}^2,
\end{equation}

\noindent where $x_i(i= 1, 2, 3)$ are the orthogonal spatial coordinates for the boundary. We assume the background is isotropic $G_{x_1 x_1} = G_{x_2 x_2} = G_{x_3 x_3}$, in which the thermal properties of static heavy $Q\bar{Q}$  pair are studied in Refs.~\cite{Kinar:1998vq,Jo:2009xr,Chamblin:1999hg}.

We choose a reference frame where the plasma is at rest, in which the $Q\bar{Q}$ dipole is moving with a constant rapidity $\eta$ along the $x_{3}$ direction. Equally, we can boost to a reference frame where the dipole is at rest but the plasma is moving past it.
Then, let us boost our reference frame in the $x_{3}$ direction with rapidity $\eta$, so that the $Q\bar{Q}$ is now at rest and the plasma moves with rapidity $-\eta$ in the $x_{3}$ direction. In our discussion, the $Q\bar{Q}$ now is at rest and feels a hot plasma wind,
\begin{gather}
\dd{t} = \dd{t}' \cosh \eta - \dd{x_3}' \sinh \eta \\
\dd{x_3} = -\dd{t}' \sinh \eta + \dd{x_3}' \cosh \eta.
\end{gather}

After dropping the prime, we get the metric
\begin{equation}
    \begin{split}
        \dd{s}^2 =& - (G_{tt} \cosh^2\eta - G_{xx} \sinh^2 \eta ) \dd{t}^2 \\
        &+(G_{xx} \cosh^2\eta - G_{tt} \sinh^2 \eta) \dd{x_3}^2 \\
        &-2\sinh \eta \cosh \eta (G_{xx} - G_{tt}) \dd{t}\dd{x_3} \\
        &+G_{xx}(\dd{x_1}^2 + \dd{x_2}^2) + G_{zz} \dd{z}^2,
    \end{split}
\end{equation}

\noindent where $G_{xx} = \frac{R^2h(z)}{z^2}, \,G_{tt}=\frac{R^2h(z)}{z^2}f(z),\, G_{zz} = \frac{R^2h(z)}{z^2f(z)}$ in the deformed AdS/RN black hole metric. We will discuss that the moving $Q\bar{Q}$  pair is aligned perpendicularly to the plasma wind.

The Nambu-Goto action of the world sheet in the Minkowski metric:
\begin{equation}
    S_{NG} = - \frac{1}{2\pi\alpha'}\int \dd[2]{\xi} \sqrt{- \det g_{ab}},
\end{equation}
where $g_{ab}$ is the induced metric on the world sheet and $\frac{1}{2\pi\alpha'}$ is the string tension, and
\begin{equation}
    g_{ab} = g_{MN} \partial_a X^M \partial_b X^N, \quad a,\,b=0,\,1,
\end{equation}
where $X^M$ and $g_{MN}$ are the coordinates and the metric of the AdS space. We use the static gauge $\xi^0 = t,\,\xi^1 = x_1$. The Nambu-Goto action is given as,
\begin{equation}
    S_{NG} = - \frac{R^2}{2\pi\alpha'T}\int_{-L/2}^{L/2} \dd{x_1} \sqrt{g_1(z) \frac{\dd{z}^2}{\dd{x}^2} + g_2(z)},
\end{equation}
where
\begin{gather}
    g_1(z) = a_1(z) \cosh^2 \eta - b_1(z)\sinh^2 \eta \notag \\
    g_2(z) = a_2(z) \cosh^2 \eta - b_2(z)\sinh^2 \eta,
\end{gather}
and
\begin{equation}
    a_1(z)=G_{tt}G_{zz},\,a_2(z)=G_{tt}G_{xx},\,b_1(z)=G_{xx}G_{zz},\,b_2(z)=G_{xx}^2.
\end{equation}

The separating distances of $Q\bar{Q}$  pair is
\begin{equation}
    L = 2 \int_0^{z_0} \qty[\frac{g_2(z)}{g_1(z)} \qty(\frac{g_2(z)}{g_2(z_0)}-1)]^{-1/2} \dd{z}.
\end{equation}

The free energy of $Q\bar{Q}$  pair is given as,
\begin{equation}
    \frac{\pi F_{Q\bar{Q}}}{\sqrt{\lambda}} = \int_0^{z_0} \dd{z}\qty(\sqrt{\frac{g_2(z)g_1(z)}{g_2(z)-g_2(z_0)}} - \sqrt{g_2(z\rightarrow 0)}) - \int_{z_0}^\infty \sqrt{g_2(z\rightarrow 0)} \dd{z}.
\end{equation}

The concept of effective string tension was first proposed in Ref.~\cite{Andreev:2006eh}, which was mainly used to discuss the thermal phase transition characteristics of a static $Q\bar{Q}$  pair. The effective string tension of the moving quarkonium can be given as,
\begin{equation}
    \sigma(z) = \sqrt{g_2(z)} = \frac{h}{z^2}\sqrt{f(z)\cosh^2 \eta - \sinh^2 \eta}.
\end{equation}

When considering the behavior of a string bit, we have the effective potential $V=\sigma(z)$. The entropy of $Q\bar{Q}$  pair can be calculated by
\begin{equation}
    S_{Q\bar{Q}} = - \partial F_{Q\bar{Q}} / \partial T,
\end{equation}
where $T$ is the temperature of the plasma.

\section{Effective String Tension and Phase Diagram}\label{sec:03}

It is well known that QCD is in the deconfinement phase at a high temperature and large chemical potential, while it is in the confinement phase at a low temperature and small chemical potential. The study of the phase structure of QCD is an important and challenging assignment. It is generally believed that there is a phase transition between the two phases. How to get phase diagrams in the $\mu$--$T$ plane is a rather difficult job because the QCD coupling constant becomes very large near the phase change region, and the traditional perturbation QCD method can not be used. For a long time, the lattice QCD method is considered as the only way to solve the problem. Although lattice QCD works well for zero density, it encounters the sign problem when considering finite density, i.e., $\mu \neq 0$. However, the most interesting region in the QCD phase diagram is at a finite density. The most concerned subjects, such as heavy-ion collisions and compact stars in astrophysics, are all related to QCD at a finite density.

This situation has greatly improved with the advent of the AdS/CFT correspondence that revived interest in finding a string description of strong interactions. Its evolving theory called AdS/QCD uses a five-dimensional effective description and attempts to fit QCD as much as possible. The effective string tension and thermal phase transition based on AdS/QCD theory are explored  in this paper.

\begin{figure}
    \centering
    \includegraphics[width=8.5cm]{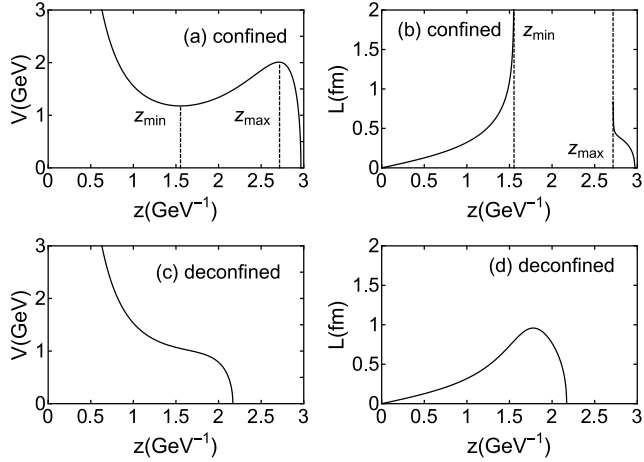}
    \caption{\label{fig3}The effective potential corresponds to the confined situation(a) and deconfined phase(c), respectively. The dependences of interquark distances $L(z)$ on the fifth holographic coordinate $z$ are shown in the confined(b) and deconfined phase(d), respectively.}
\end{figure}

In Eq.(16), we give the effective potential $V$ of the moving quarkonium. It is found that the effective string potential $V$ is a function of temperature, chemical potential and rapidity.
For fixed values of the chemical potential $\mu = 0.1\,\mathrm{GeV}$, rapidity $\eta = 0.3$ and temperature $T = 0.1\,\mathrm{GeV}$, the dependence of the effective potential $V$ on the fifth holographic coordinate $z$ in the confined situation is given in Fig.~\ref{fig3}(a). The effective potential $V$ reaches a minimum when $z = z_\text{min}$, and $V$  reaches the maximum value when $z = z_\text{max}$. The dependence of distance $L(z)$ on $z$ in the confined situation is calculated from the Eq.(14) as shown in Fig.~\ref{fig3}(b). The distance $L$ increases monotonically from $L(0)=0$ and diverges when  $z \rightarrow z_\text{min}$ at $0 \leq z \leq z_\text{min}$.  But $L$ monotonically decreases from a finite value $L(z_\text{max})$ to $L(z_h) = 0$ at $z_\text{max} \leq z \leq z_h$.

The two points $z_\text{min}$ and $z_\text{max}$ get more and more close along the $z$ direction with the increase of temperature $T$. When the temperature $T$ increases to a certain value such as a critical temperature $T_{c}$,
the two points $z_\text{min}$ and $z_\text{max}$ coincide at $z_\text{min} = z_\text{max} = z_m$. Therefore, taking $\mu = 0.1\,\mathrm{GeV}$ and $T= 0.14\,\mathrm{GeV}$ $>T_c$, the effective potential $V$ can be evaluated  in Fig.~\ref{fig3}(c).  For $T \geq T_c$, $L(z)$ is real when $0 \leq z \leq z_h$ and never exceeds the value $L(z_m)$, as shown in Fig.~\ref{fig3}(d). At the point $(\mu_c, T_c)$ in the $\mu$--$T$ plane, $L(z)$ is always limited.

For $L > L(z_m)$,two strings extend between the boundary $z = 0$ and the black hole horizon $z=z_h$ as shown in Fig.~\ref{fig1}(b). The quark-antiquark pair becomes two deconfined quarks.

\begin{figure}
    \centering
    \includegraphics[width=8.5cm]{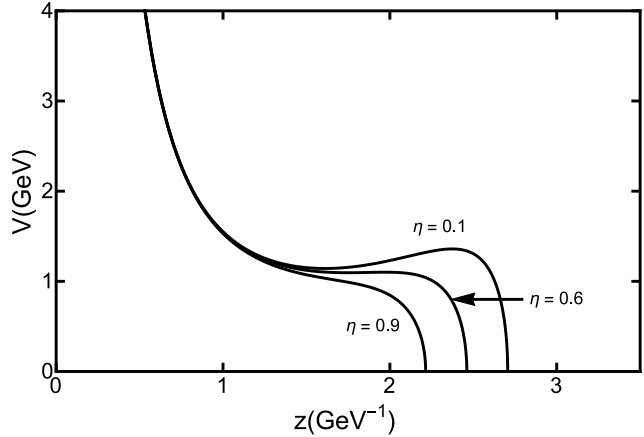}
    \caption{\label{fig4}The effective potential vs $z$ at different rapidities for a given temperature and chemical potential. It indicates the phase transition from a confinement to deconfiment phase with the variation of rapidity.}
\end{figure}

Figure~\ref{fig4} shows the dependence of the effective string tensions on $z$ at different rapidities of a moving quarkonium at a given chemical potential $\mu = 0.1\,\mathrm{GeV}$ and a temperature $T = 0.1\,\mathrm{GeV}$. It is found that when $\eta = 0.1$ and $z = z_\text{min}$, $V$ reaches a minimum, and when $\eta = 0.1$ and $z = z_\text{max}$, $V$ reaches the maximum value. The two points $z_\text{min}$ and $z_\text{max}$ along the $z$ axis get more and more close with the increase of the rapidity $\eta$.  When the rapidity $\eta$ increases to a certain value, the two points $z_\text{min}$ and $z_\text{max}$ coincide at $z_\text{min} = z_\text{max} = z_m$, which means the transition from a confinement to deconfinement phase at the given temperature and rapidity.

\begin{figure}
    \centering
    \includegraphics[width=8.5cm]{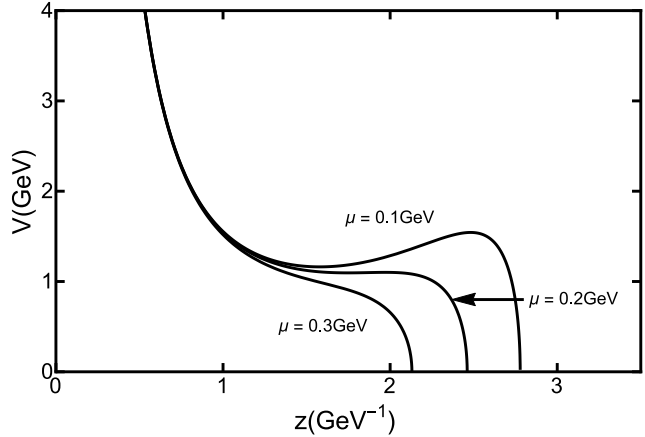}
    \caption{\label{fig5}The effective potential vs $z$ at different chemical potentials for a given temperature and rapidity. It indicates the phase transition from a confinement to deconfinement phase with the variation of chemical potential.}
\end{figure}

The dependence of the effective string tensions on $z$ at different chemical potentials $\mu$ under a given rapidity $\eta = 0.3$ and a temperature $T = 0.1\, \mathrm{GeV}$ is shown in Fig.~\ref{fig5}. One finds that when $\mu = 0.1\,\mathrm{GeV}$ and $z = z_\text{min}$, $V$ gets a minimum, and when $\mu = 0.1\,\mathrm{GeV}$ and $z = z_\text{max}$, $V$ gets the maximum value.
 When the chemical potential $\mu$ increases to a certain value such as the critical chemical potential $\mu_{c}$, the two points $z_\text{min}$ and $z_\text{max}$ coincide at $z_\text{min} = z_\text{max} = z_m$,  which corresponds to the change from a confinement to deconfinement phase under a given temperature and rapidity.

In view of the above analysis and calculation, we can obtain the phase diagram in the $\mu$--$T$  plane. The curve defined by the points $[\mu, T_{c}(\mu)]$  is shown in Fig. 6. This can be regarded as the deconfinement transition. The picture agrees with the diagram obtained by, e.g, Nambu-Jona-Lasinio\cite{Buballa:2003qv} and other effective models~\cite{Schaefer:2004en}. The physical picture can be understood as a static probe put in the expansive system. that is to say, the system moves through the probe with a certain rapidity.
From our study, we argue that the moving system will change the feature of the phase transition.
We find that a moving system will reach the phase transition point at a lower temperature and chemical potential than a stationary system. It means that the lifetime of the moving QGP become longer than the static one.

\begin{figure}
    \centering
    \includegraphics[width=8.5cm]{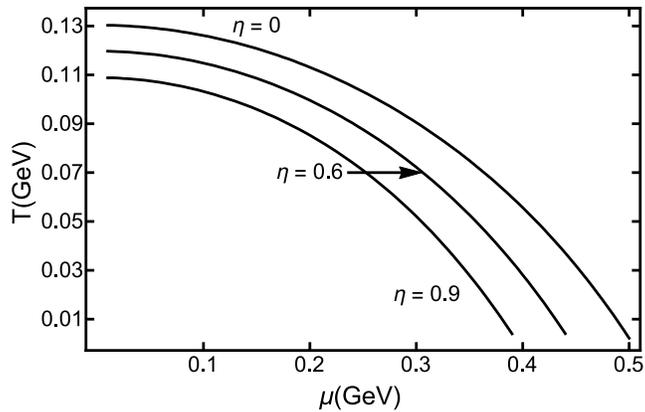}
    \caption{\label{fig6} The phase diagram in the $\mu$--$T$  plane at different rapidities of the moving system.}
\end{figure}

\begin{figure}
    \centering
    \includegraphics[width=8.5cm]{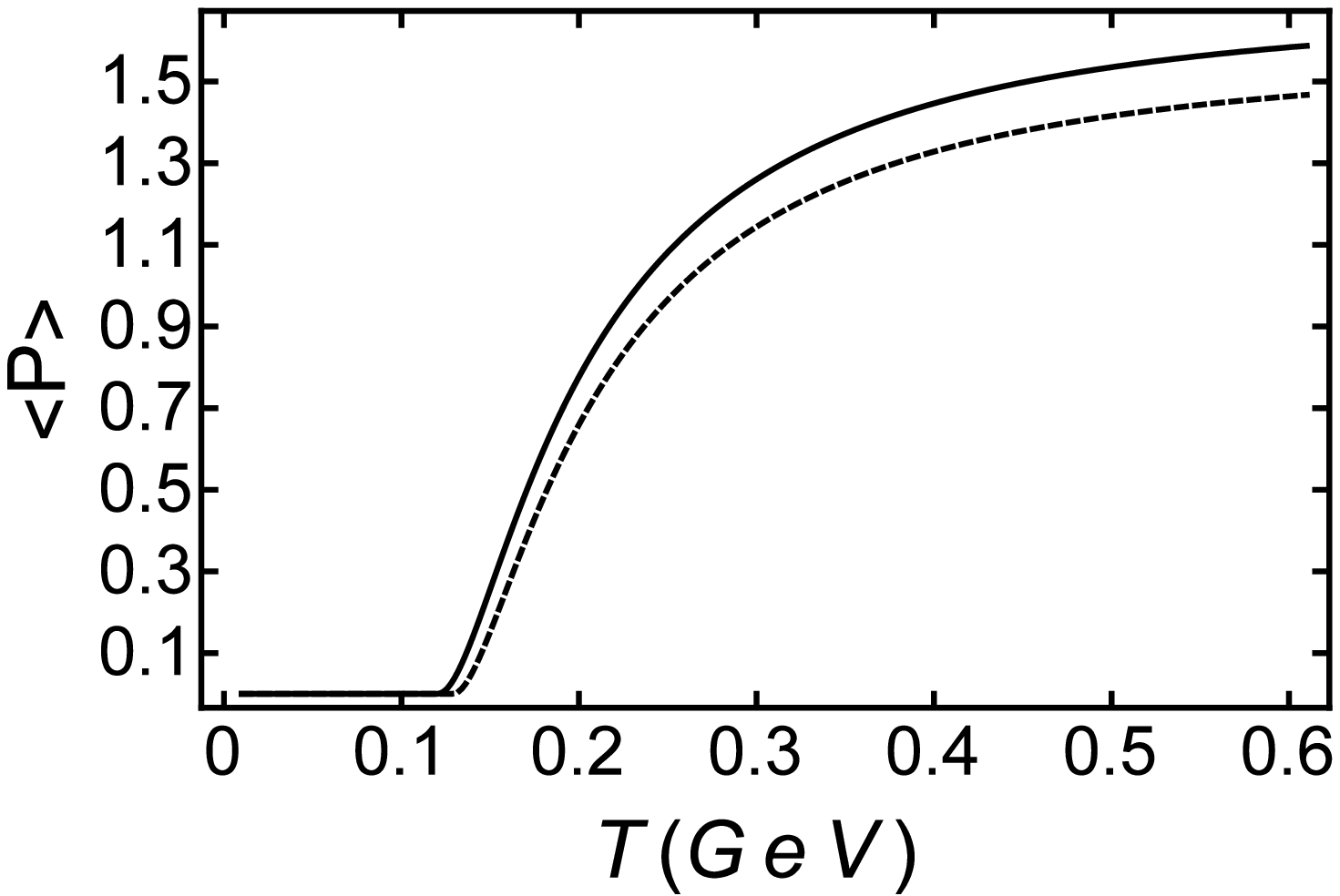}
    \caption{\label{fig7}
    Polyakov loop $<P>$ for $\mu = 0.1\,\mathrm{GeV}$, $\eta = 0$ (dashed line)and  $\mu = 0.1\,\mathrm{GeV}$, $\eta = 0.6$(solid line).}
\end{figure}

\begin{figure}
    \centering
    \includegraphics[width=8.5cm]{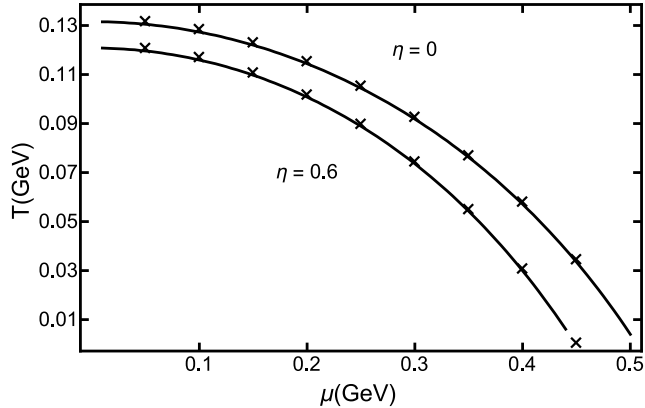}
    \caption{\label{fig8} The solid line represents the results calculated from effective string tension and crosses denote the results calculated from a Polyakov loop at different rapidities.}
\end{figure}

A Polyakov loop is an order parameter of the confinement-deconfinement transition often used in many articles~\cite{Cai:2012xh,Colangelo:2010pe,Andreev:2006nw}. But we find that the effective string tension is also a useful physical quantity to study the confinement-deconfinement transition in holographic QCD. As a comparison, we add the Polyakov loop in our article shown in Fig.~\ref{fig7}. Figure~\ref{fig8} provides the comparison between a Polyakov loop and effective string tension. It is found that the results from a Polyakov loop are in good agreement  with the phase diagram calculated by using the effective string tension. The definition of a Polyakov loop can be found in Ref.~\cite{Colangelo:2010pe}. The effective string tension turns out to be useful in studying symmetry breaking(phase transition) within field theories at a finite temperature~\cite{Andreev:2006nw}. We extend it to a finite chemical potential and rapidity and testify its validity.

\section{THE ENTROPY OF A MOVING QUARKONIUM IN DEFORMED AdS/REISSNER-NORDSTROM THEORY}\label{sec:04}

\begin{figure}
    \centering
    \includegraphics[width=8.5cm]{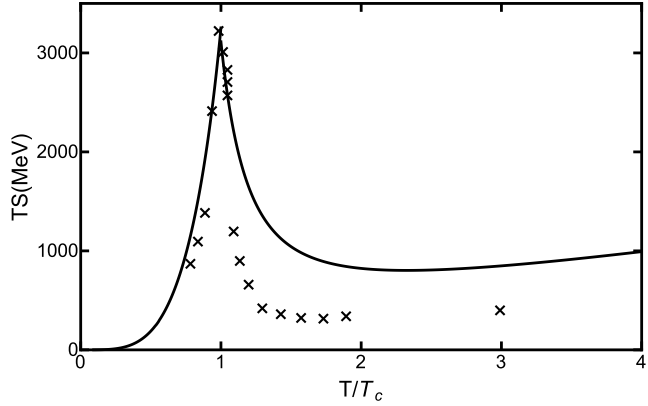}
    \caption{\label{fig9}A comparison between our calculated results and the lattice QCD results. The solid line is our calculated results, and the crosses are the lattice QCD results.}
\end{figure}

The lattice QCD shows  the presence of an additional entropy associated with a static heavy $Q\bar{Q}$  pair in the QCD plasma, and there is a peak of entropy at the deconfinement transition\cite{Kaczmarek:2002mc,Petreczky:2004pz,Kaczmarek:2005zp} as shown in Fig.~\ref{fig9}. We calculate the entropy of a static heavy $Q\bar{Q}$  pair at a chemical potential  $\mu \rightarrow 0$ in the soft wall model. We use Eq.(17)to calculate the entropy of quark-antiquark pair below the critical temperature. Above the critical temperature, the entropy of the pair equals the sum of the entropy of two single quarks. The entropy of a single quark can be calculated by Eq.(17), but $z$ is on the interval [0,$z_{h}$]. We fix $\lambda = 2.6$ in our calculation.

It is found that our calculated results show a peak of the heavy quark-antiquark entropy in the deconfinement transition and fits well with the lattice QCD theory. Comparing with the SYM theory\cite{Hashimoto:2014fha}, we find the necessity of the deformed metric in our study.

To study the entropic destruction of the moving quarkonium, Ref.~\cite{Fadafan:2015ynz} calculated the entropy of a moving quarkonium by using $N=4$ SYM theory and considered the quarkonium moves parallel ($\theta = 0$) and perpendicularly($\theta = \pi/2$) to the wind. We calculate the entropy in the transverse direction ($\theta = \pi/2$) by considering the deformed metric.

\begin{figure}
    \centering
    \includegraphics[width=8.5cm]{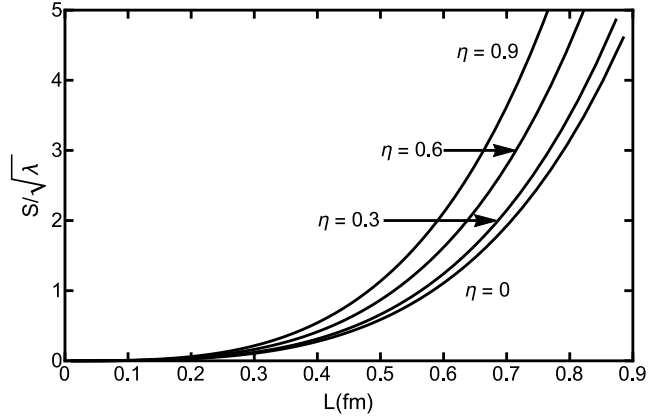}
    \caption{\label{fig10}The entropy of a moving quarkonium as a function of $L$ at different rapidities. We fix the chemical potential as $\mu = 0.1\,\mathrm{GeV}$ and the temperature as $T = 0.1\,\mathrm{GeV}$.}
\end{figure}

\begin{figure}
    \centering
    \includegraphics[width=8.5cm]{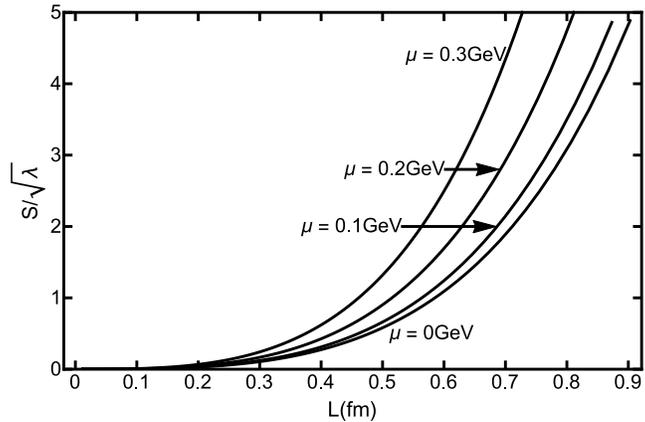}
    \caption{\label{fig11}The entropy of a moving quarkonium as a function of $L$ at different chemical potentials. We fix the rapidity of a moving quarkonium as $\eta = 0.3$ and the temperature as $T = 0.1\,\mathrm{GeV}$.}
\end{figure}

Figure~\ref{fig10} shows the dependence of confined heavy-quarkonium entropy on the interquark distance $L$ at different rapidities. We fix the chemical potential as $\mu = 0.1\,\mathrm{GeV}$  and the temperature as $T = 0.1\,\mathrm{GeV}$. As it is shown in this figure, increasing the velocity of quarkonium leads to stronger entropy at small distances. We can conclude that the quarkonium will be easier to dissociate by increasing the velocity of quarkonium.

Figure~\ref{fig11} shows the dependence of confined heavy-quarkonium entropy on the interquark distance $L$ at different chemical potentials. We fix the rapidity as $\eta = 0.3$ and the temperature as $T = 0.1\,\mathrm{GeV}$. As it
is shown in this figure, increasing the chemical potential leads to stronger entropy at small distances. We also conclude that the quarkonium will be easier to dissociate by increasing the chemical potential.

\section{Summary and Conclusions}\label{sec:05}
Lattice QCD predicts that the heavy $Q\bar{Q}$  pair immersed in the QGP possess a large amount of entropy. This indicates a strong degree of entanglement between the pair and the rest of the system. The sharp peak of the entropy at the deconfinement transition is a salient feature in the lattice data. It is
very likely that the quark-gluon plasma around the deconfinement transition is strongly coupled; therefore, in this paper, we have used the deformed metric to study the entropy associated with the heavy $Q\bar{Q}$  pair. We show that the quarkonium will be easier to dissociate  by increasing the velocity of a quarkonium or chemical potential.

It is a standard procedure to study a confinement-deconfinement phase transition
by checking the different configurations of a quark and antiquark in AdS/QCD~\cite{Cai:2012xh,Colangelo:2010pe,Yang:2015aia}.
The heavy quark pair can be regarded as a probe to detect whether the system is in a confinement or deconfinement phase.
The moving case of the quark-antiquark pair has been calculated in many aspects~\cite{Finazzo:2014rca,Liu:2006nn,Avramis:2006em}. As mentioned in Ref.~\cite{Finazzo:2014rca}, you can choose a reference frame where the plasma is at rest and the $Q\bar{Q}$ dipole is moving with a constant velocity. Equally, you can boost to a reference frame where the $Q\bar{Q}$ dipole is at rest but the plasma is moving past it.
The $Q\bar{Q}$ pair is now at rest and feels a hot plasma wind in our discussion. The physical picture can be understood as a static probe located in the expansive system. That is, the system moves through the probe with a certain rapidity. From our study, we argue that a moving system will change the feature of the phase transition.

We find that a moving system will reach the phase transition point at a lower temperature or at a smaller chemical potential than a stationary system. It means that the lifetime of the moving QGP become longer than the static one. It is important to study the phase transition of a moving plasma system because the QCD system produced in relativistic heavy ion collisions is not in a static state. Studying the phase transition of a moving system will help us to understand the mechanism of a QGP phase transition in relativistic heavy ion collisions.

By analyzing the characteristics of the effective string tension of a heavy quark-antiquark pair with different temperatures, chemical potentials, and rapidities, we systematically study the feature of the confinement-deconfinement phase transition. It is found that the confinement-deconfinement transition detected by the behavior of the quarkonium occurs at a lower temperature or at a smaller chemical potential, when the media is boosted.

\section*{Acknowledgments}
This work was supported by National Natural Science Foundation of China (Grant No. 11747115,No. 11475068), the CCNU-QLPL Innovation Fund (Grant
No. QLPL2016P01), the Excellent Youth Foundation of Hubei Scientific Committee (Grant No. 2006ABB036).

\section*{References}

\bibliography{ref}

\begin{thebibliography}{42}%
\makeatletter
\providecommand \@ifxundefined [1]{%
 \@ifx{#1\undefined}
}%
\providecommand \@ifnum [1]{%
 \ifnum #1\expandafter \@firstoftwo
 \else \expandafter \@secondoftwo
 \fi
}%
\providecommand \@ifx [1]{%
 \ifx #1\expandafter \@firstoftwo
 \else \expandafter \@secondoftwo
 \fi
}%
\providecommand \natexlab [1]{#1}%
\providecommand \enquote  [1]{``#1''}%
\providecommand \bibnamefont  [1]{#1}%
\providecommand \bibfnamefont [1]{#1}%
\providecommand \citenamefont [1]{#1}%
\providecommand \href@noop [0]{\@secondoftwo}%
\providecommand \href [0]{\begingroup \@sanitize@url \@href}%
\providecommand \@href[1]{\@@startlink{#1}\@@href}%
\providecommand \@@href[1]{\endgroup#1\@@endlink}%
\providecommand \@sanitize@url [0]{\catcode `\\12\catcode `\$12\catcode
  `\&12\catcode `\#12\catcode `\^12\catcode `\_12\catcode `\%12\relax}%
\providecommand \@@startlink[1]{}%
\providecommand \@@endlink[0]{}%
\providecommand \url  [0]{\begingroup\@sanitize@url \@url }%
\providecommand \@url [1]{\endgroup\@href {#1}{\urlprefix }}%
\providecommand \urlprefix  [0]{URL }%
\providecommand \Eprint [0]{\href }%
\providecommand \doibase [0]{http://dx.doi.org/}%
\providecommand \selectlanguage [0]{\@gobble}%
\providecommand \bibinfo  [0]{\@secondoftwo}%
\providecommand \bibfield  [0]{\@secondoftwo}%
\providecommand \translation [1]{[#1]}%
\providecommand \BibitemOpen [0]{}%
\providecommand \bibitemStop [0]{}%
\providecommand \bibitemNoStop [0]{.\EOS\space}%
\providecommand \EOS [0]{\spacefactor3000\relax}%
\providecommand \BibitemShut  [1]{\csname bibitem#1\endcsname}%
\let\auto@bib@innerbib\@empty
\bibitem [{\citenamefont {He}\ \emph {et~al.}(2010)\citenamefont {He},
  \citenamefont {Huang}, \citenamefont {Yan},\ and\ \citenamefont
  {Yang}}]{Huang:2007fv}%
  \BibitemOpen
  \bibfield  {author} {\bibinfo {author} {\bibfnamefont {S.}~\bibnamefont
  {He}}, \bibinfo {author} {\bibfnamefont {M.}~\bibnamefont {Huang}}, \bibinfo
  {author} {\bibfnamefont {Q.-S.}\ \bibnamefont {Yan}}, \ and\ \bibinfo
  {author} {\bibfnamefont {Y.}~\bibnamefont {Yang}},\ }\href {\doibase
  10.1140/epjc/s10052-010-1239-0} {\bibfield  {journal} {\bibinfo  {journal}
  {Eur. Phys. J. C}\ }\textbf {\bibinfo {volume} {66}},\ \bibinfo {pages} {187}
  (\bibinfo {year} {2010})}\BibitemShut {NoStop}%
\bibitem [{\citenamefont {Kruczenski}\ \emph {et~al.}(2004)\citenamefont
  {Kruczenski}, \citenamefont {Mateos}, \citenamefont {Myers},\ and\
  \citenamefont {Winters}}]{Kruczenski:2003uq}%
  \BibitemOpen
  \bibfield  {author} {\bibinfo {author} {\bibfnamefont {M.}~\bibnamefont
  {Kruczenski}}, \bibinfo {author} {\bibfnamefont {D.}~\bibnamefont {Mateos}},
  \bibinfo {author} {\bibfnamefont {R.~C.}\ \bibnamefont {Myers}}, \ and\
  \bibinfo {author} {\bibfnamefont {D.~J.}\ \bibnamefont {Winters}},\ }\href
  {\doibase 10.1088/1126-6708/2004/05/041} {\bibfield  {journal} {\bibinfo
  {journal} {JHEP}\ }\textbf {\bibinfo {volume} {05}},\ \bibinfo {pages} {041}
  (\bibinfo {year} {2004})}\BibitemShut {NoStop}%
\bibitem [{\citenamefont {Sakai}\ and\ \citenamefont
  {Sugimoto}(2005{\natexlab{a}})}]{Sakai:2004cn}%
  \BibitemOpen
  \bibfield  {author} {\bibinfo {author} {\bibfnamefont {T.}~\bibnamefont
  {Sakai}}\ and\ \bibinfo {author} {\bibfnamefont {S.}~\bibnamefont
  {Sugimoto}},\ }\href {\doibase 10.1143/PTP.113.843} {\bibfield  {journal}
  {\bibinfo  {journal} {Prog. Theor. Phys.}\ }\textbf {\bibinfo {volume}
  {113}},\ \bibinfo {pages} {843} (\bibinfo {year}
  {2005}{\natexlab{a}})}\BibitemShut {NoStop}%
\bibitem [{\citenamefont {Sakai}\ and\ \citenamefont
  {Sugimoto}(2005{\natexlab{b}})}]{Sakai:2005yt}%
  \BibitemOpen
  \bibfield  {author} {\bibinfo {author} {\bibfnamefont {T.}~\bibnamefont
  {Sakai}}\ and\ \bibinfo {author} {\bibfnamefont {S.}~\bibnamefont
  {Sugimoto}},\ }\href {\doibase 10.1143/PTP.114.1083} {\bibfield  {journal}
  {\bibinfo  {journal} {Prog. Theor. Phys.}\ }\textbf {\bibinfo {volume}
  {114}},\ \bibinfo {pages} {1083} (\bibinfo {year}
  {2005}{\natexlab{b}})}\BibitemShut {NoStop}%
\bibitem [{\citenamefont {Witten}(1998)}]{Witten:1998zw}%
  \BibitemOpen
  \bibfield  {author} {\bibinfo {author} {\bibfnamefont {E.}~\bibnamefont
  {Witten}},\ }\href@noop {} {\bibfield  {journal} {\bibinfo  {journal} {Adv.
  Theor. Math. Phys.}\ }\textbf {\bibinfo {volume} {2}},\ \bibinfo {pages}
  {505} (\bibinfo {year} {1998})}\BibitemShut {NoStop}%
\bibitem [{\citenamefont {Brandhuber}\ \emph {et~al.}(1998)\citenamefont
  {Brandhuber}, \citenamefont {Itzhaki}, \citenamefont {Sonnenschein},\ and\
  \citenamefont {Yankielowicz}}]{Brandhuber:1998bs}%
  \BibitemOpen
  \bibfield  {author} {\bibinfo {author} {\bibfnamefont {A.}~\bibnamefont
  {Brandhuber}}, \bibinfo {author} {\bibfnamefont {N.}~\bibnamefont {Itzhaki}},
  \bibinfo {author} {\bibfnamefont {J.}~\bibnamefont {Sonnenschein}}, \ and\
  \bibinfo {author} {\bibfnamefont {S.}~\bibnamefont {Yankielowicz}},\ }\href
  {\doibase 10.1016/S0370-2693(98)00730-8} {\bibfield  {journal} {\bibinfo
  {journal} {Phys. Lett.}\ }\textbf {\bibinfo {volume} {B434}},\ \bibinfo
  {pages} {36} (\bibinfo {year} {1998})}\BibitemShut {NoStop}%
\bibitem [{\citenamefont {Rey}\ \emph {et~al.}(1998)\citenamefont {Rey},
  \citenamefont {Theisen},\ and\ \citenamefont {Yee}}]{Rey:1998bq}%
  \BibitemOpen
  \bibfield  {author} {\bibinfo {author} {\bibfnamefont {S.-J.}\ \bibnamefont
  {Rey}}, \bibinfo {author} {\bibfnamefont {S.}~\bibnamefont {Theisen}}, \ and\
  \bibinfo {author} {\bibfnamefont {J.-T.}\ \bibnamefont {Yee}},\ }\href
  {\doibase 10.1016/S0550-3213(98)00471-4} {\bibfield  {journal} {\bibinfo
  {journal} {Nucl. Phys.}\ }\textbf {\bibinfo {volume} {B527}},\ \bibinfo
  {pages} {171} (\bibinfo {year} {1998})}\BibitemShut {NoStop}%
\bibitem [{\citenamefont {Kinar}\ \emph {et~al.}(2000)\citenamefont {Kinar},
  \citenamefont {Schreiber},\ and\ \citenamefont
  {Sonnenschein}}]{Kinar:1998vq}%
  \BibitemOpen
  \bibfield  {author} {\bibinfo {author} {\bibfnamefont {Y.}~\bibnamefont
  {Kinar}}, \bibinfo {author} {\bibfnamefont {E.}~\bibnamefont {Schreiber}}, \
  and\ \bibinfo {author} {\bibfnamefont {J.}~\bibnamefont {Sonnenschein}},\
  }\href {\doibase 10.1016/S0550-3213(99)00652-5} {\bibfield  {journal}
  {\bibinfo  {journal} {Nucl. Phys.}\ }\textbf {\bibinfo {volume} {B566}},\
  \bibinfo {pages} {103} (\bibinfo {year} {2000})}\BibitemShut {NoStop}%
\bibitem [{\citenamefont {He}\ \emph {et~al.}(2011)\citenamefont {He},
  \citenamefont {Huang},\ and\ \citenamefont {Yan}}]{He:2010ye}%
  \BibitemOpen
  \bibfield  {author} {\bibinfo {author} {\bibfnamefont {S.}~\bibnamefont
  {He}}, \bibinfo {author} {\bibfnamefont {M.}~\bibnamefont {Huang}}, \ and\
  \bibinfo {author} {\bibfnamefont {Q.-S.}\ \bibnamefont {Yan}},\ }\href
  {\doibase 10.1103/PhysRevD.83.045034} {\bibfield  {journal} {\bibinfo
  {journal} {Phys. Rev.}\ }\textbf {\bibinfo {volume} {D83}},\ \bibinfo {pages}
  {045034} (\bibinfo {year} {2011})}\BibitemShut {NoStop}%
\bibitem [{\citenamefont {Pirner}\ and\ \citenamefont
  {Galow}(2009)}]{Pirner:2009gr}%
  \BibitemOpen
  \bibfield  {author} {\bibinfo {author} {\bibfnamefont {H.~J.}\ \bibnamefont
  {Pirner}}\ and\ \bibinfo {author} {\bibfnamefont {B.}~\bibnamefont {Galow}},\
  }\href {\doibase 10.1016/j.physletb.2009.07.009} {\bibfield  {journal}
  {\bibinfo  {journal} {Phys. Lett.}\ }\textbf {\bibinfo {volume} {B679}},\
  \bibinfo {pages} {51} (\bibinfo {year} {2009})}\BibitemShut {NoStop}%
\bibitem [{\citenamefont {Shock}\ \emph {et~al.}(2007)\citenamefont {Shock},
  \citenamefont {Wu}, \citenamefont {Wu},\ and\ \citenamefont
  {Xie}}]{Shock:2006gt}%
  \BibitemOpen
  \bibfield  {author} {\bibinfo {author} {\bibfnamefont {J.~P.}\ \bibnamefont
  {Shock}}, \bibinfo {author} {\bibfnamefont {F.}~\bibnamefont {Wu}}, \bibinfo
  {author} {\bibfnamefont {Y.-L.}\ \bibnamefont {Wu}}, \ and\ \bibinfo {author}
  {\bibfnamefont {Z.-F.}\ \bibnamefont {Xie}},\ }\href {\doibase
  10.1088/1126-6708/2007/03/064} {\bibfield  {journal} {\bibinfo  {journal}
  {JHEP}\ }\textbf {\bibinfo {volume} {03}},\ \bibinfo {pages} {064} (\bibinfo
  {year} {2007})}\BibitemShut {NoStop}%
\bibitem [{\citenamefont {Zeng}(2008)}]{Zeng:2008sx}%
  \BibitemOpen
  \bibfield  {author} {\bibinfo {author} {\bibfnamefont {D.-f.}\ \bibnamefont
  {Zeng}},\ }\href {\doibase 10.1103/PhysRevD.78.126006} {\bibfield  {journal}
  {\bibinfo  {journal} {Phys. Rev.}\ }\textbf {\bibinfo {volume} {D78}},\
  \bibinfo {pages} {126006} (\bibinfo {year} {2008})}\BibitemShut {NoStop}%
\bibitem [{\citenamefont {Brodsky}\ \emph {et~al.}(2010)\citenamefont
  {Brodsky}, \citenamefont {de~Teramond},\ and\ \citenamefont
  {Deur}}]{Brodsky:2010ur}%
  \BibitemOpen
  \bibfield  {author} {\bibinfo {author} {\bibfnamefont {S.~J.}\ \bibnamefont
  {Brodsky}}, \bibinfo {author} {\bibfnamefont {G.~F.}\ \bibnamefont
  {de~Teramond}}, \ and\ \bibinfo {author} {\bibfnamefont {A.}~\bibnamefont
  {Deur}},\ }\href {\doibase 10.1103/PhysRevD.81.096010} {\bibfield  {journal}
  {\bibinfo  {journal} {Phys. Rev.}\ }\textbf {\bibinfo {volume} {D81}},\
  \bibinfo {pages} {096010} (\bibinfo {year} {2010})}\BibitemShut {NoStop}%
\bibitem [{\citenamefont {Ghoroku}\ \emph {et~al.}(2003)\citenamefont
  {Ghoroku}, \citenamefont {Tachibana},\ and\ \citenamefont
  {Uekusa}}]{Ghoroku:2003ex}%
  \BibitemOpen
  \bibfield  {author} {\bibinfo {author} {\bibfnamefont {K.}~\bibnamefont
  {Ghoroku}}, \bibinfo {author} {\bibfnamefont {M.}~\bibnamefont {Tachibana}},
  \ and\ \bibinfo {author} {\bibfnamefont {N.}~\bibnamefont {Uekusa}},\ }\href
  {\doibase 10.1103/PhysRevD.68.125002} {\bibfield  {journal} {\bibinfo
  {journal} {Phys. Rev.}\ }\textbf {\bibinfo {volume} {D68}},\ \bibinfo {pages}
  {125002} (\bibinfo {year} {2003})}\BibitemShut {NoStop}%
\bibitem [{\citenamefont {Gursoy}\ \emph {et~al.}(2008)\citenamefont {Gursoy},
  \citenamefont {Kiritsis},\ and\ \citenamefont {Nitti}}]{Gursoy:2007er}%
  \BibitemOpen
  \bibfield  {author} {\bibinfo {author} {\bibfnamefont {U.}~\bibnamefont
  {Gursoy}}, \bibinfo {author} {\bibfnamefont {E.}~\bibnamefont {Kiritsis}}, \
  and\ \bibinfo {author} {\bibfnamefont {F.}~\bibnamefont {Nitti}},\ }\href
  {\doibase 10.1088/1126-6708/2008/02/019} {\bibfield  {journal} {\bibinfo
  {journal} {JHEP}\ }\textbf {\bibinfo {volume} {02}},\ \bibinfo {pages} {019}
  (\bibinfo {year} {2008})}\BibitemShut {NoStop}%
\bibitem [{\citenamefont {Maldacena}(1998)}]{Maldacena:1998im}%
  \BibitemOpen
  \bibfield  {author} {\bibinfo {author} {\bibfnamefont {J.~M.}\ \bibnamefont
  {Maldacena}},\ }\href {\doibase 10.1103/PhysRevLett.80.4859} {\bibfield
  {journal} {\bibinfo  {journal} {Phys. Rev. Lett.}\ }\textbf {\bibinfo
  {volume} {80}},\ \bibinfo {pages} {4859} (\bibinfo {year}
  {1998})}\BibitemShut {NoStop}%
\bibitem [{\citenamefont {Karch}\ \emph {et~al.}(2006)\citenamefont {Karch},
  \citenamefont {Katz}, \citenamefont {Son},\ and\ \citenamefont
  {Stephanov}}]{Karch:2006pv}%
  \BibitemOpen
  \bibfield  {author} {\bibinfo {author} {\bibfnamefont {A.}~\bibnamefont
  {Karch}}, \bibinfo {author} {\bibfnamefont {E.}~\bibnamefont {Katz}},
  \bibinfo {author} {\bibfnamefont {D.~T.}\ \bibnamefont {Son}}, \ and\
  \bibinfo {author} {\bibfnamefont {M.~A.}\ \bibnamefont {Stephanov}},\ }\href
  {\doibase 10.1103/PhysRevD.74.015005} {\bibfield  {journal} {\bibinfo
  {journal} {Phys. Rev.}\ }\textbf {\bibinfo {volume} {D74}},\ \bibinfo {pages}
  {015005} (\bibinfo {year} {2006})}\BibitemShut {NoStop}%
\bibitem [{\citenamefont {Andreev}\ and\ \citenamefont
  {Zakharov}(2007{\natexlab{a}})}]{Andreev:2006eh}%
  \BibitemOpen
  \bibfield  {author} {\bibinfo {author} {\bibfnamefont {O.}~\bibnamefont
  {Andreev}}\ and\ \bibinfo {author} {\bibfnamefont {V.~I.}\ \bibnamefont
  {Zakharov}},\ }\href {\doibase 10.1016/j.physletb.2007.01.002} {\bibfield
  {journal} {\bibinfo  {journal} {Phys. Lett.}\ }\textbf {\bibinfo {volume}
  {B645}},\ \bibinfo {pages} {437} (\bibinfo {year}
  {2007}{\natexlab{a}})}\BibitemShut {NoStop}%
\bibitem [{\citenamefont {Kajantie}\ \emph {et~al.}(2007)\citenamefont
  {Kajantie}, \citenamefont {Tahkokallio},\ and\ \citenamefont
  {Yee}}]{Kajantie:2006hv}%
  \BibitemOpen
  \bibfield  {author} {\bibinfo {author} {\bibfnamefont {K.}~\bibnamefont
  {Kajantie}}, \bibinfo {author} {\bibfnamefont {T.}~\bibnamefont
  {Tahkokallio}}, \ and\ \bibinfo {author} {\bibfnamefont {J.-T.}\ \bibnamefont
  {Yee}},\ }\href {\doibase 10.1088/1126-6708/2007/01/019} {\bibfield
  {journal} {\bibinfo  {journal} {JHEP}\ }\textbf {\bibinfo {volume} {01}},\
  \bibinfo {pages} {019} (\bibinfo {year} {2007})}\BibitemShut {NoStop}%
\bibitem [{\citenamefont {Matsui}\ and\ \citenamefont
  {Satz}(1986)}]{Matsui:1986dk}%
  \BibitemOpen
  \bibfield  {author} {\bibinfo {author} {\bibfnamefont {T.}~\bibnamefont
  {Matsui}}\ and\ \bibinfo {author} {\bibfnamefont {H.}~\bibnamefont {Satz}},\
  }\href {\doibase 10.1016/0370-2693(86)91404-8} {\bibfield  {journal}
  {\bibinfo  {journal} {Phys. Lett.}\ }\textbf {\bibinfo {volume} {B178}},\
  \bibinfo {pages} {416} (\bibinfo {year} {1986})}\BibitemShut {NoStop}%
\bibitem [{\citenamefont {Song}\ \emph {et~al.}(2008)\citenamefont {Song},
  \citenamefont {Park}, \citenamefont {Lee},\ and\ \citenamefont
  {Wong}}]{Song:2007gm}%
  \BibitemOpen
  \bibfield  {author} {\bibinfo {author} {\bibfnamefont {T.}~\bibnamefont
  {Song}}, \bibinfo {author} {\bibfnamefont {Y.}~\bibnamefont {Park}}, \bibinfo
  {author} {\bibfnamefont {S.~H.}\ \bibnamefont {Lee}}, \ and\ \bibinfo
  {author} {\bibfnamefont {C.-Y.}\ \bibnamefont {Wong}},\ }\href {\doibase
  10.1016/j.physletb.2007.11.084} {\bibfield  {journal} {\bibinfo  {journal}
  {Phys. Lett.}\ }\textbf {\bibinfo {volume} {B659}},\ \bibinfo {pages} {621}
  (\bibinfo {year} {2008})}\BibitemShut {NoStop}%
\bibitem [{\citenamefont {Escobedo}\ \emph {et~al.}(2013)\citenamefont
  {Escobedo}, \citenamefont {Giannuzzi}, \citenamefont {Mannarelli},\ and\
  \citenamefont {Soto}}]{Escobedo:2013tca}%
  \BibitemOpen
  \bibfield  {author} {\bibinfo {author} {\bibfnamefont {M.~A.}\ \bibnamefont
  {Escobedo}}, \bibinfo {author} {\bibfnamefont {F.}~\bibnamefont {Giannuzzi}},
  \bibinfo {author} {\bibfnamefont {M.}~\bibnamefont {Mannarelli}}, \ and\
  \bibinfo {author} {\bibfnamefont {J.}~\bibnamefont {Soto}},\ }\href {\doibase
  10.1103/PhysRevD.87.114005} {\bibfield  {journal} {\bibinfo  {journal} {Phys.
  Rev.}\ }\textbf {\bibinfo {volume} {D87}},\ \bibinfo {pages} {114005}
  (\bibinfo {year} {2013})}\BibitemShut {NoStop}%
\bibitem [{\citenamefont {Jo}\ \emph {et~al.}(2010)\citenamefont {Jo},
  \citenamefont {Lee}, \citenamefont {Park},\ and\ \citenamefont
  {Sin}}]{Jo:2009xr}%
  \BibitemOpen
  \bibfield  {author} {\bibinfo {author} {\bibfnamefont {K.}~\bibnamefont
  {Jo}}, \bibinfo {author} {\bibfnamefont {B.-H.}\ \bibnamefont {Lee}},
  \bibinfo {author} {\bibfnamefont {C.}~\bibnamefont {Park}}, \ and\ \bibinfo
  {author} {\bibfnamefont {S.-J.}\ \bibnamefont {Sin}},\ }\href {\doibase
  10.1007/JHEP06(2010)022} {\bibfield  {journal} {\bibinfo  {journal} {JHEP}\
  }\textbf {\bibinfo {volume} {06}},\ \bibinfo {pages} {022} (\bibinfo {year}
  {2010})}\BibitemShut {NoStop}%
\bibitem [{\citenamefont {Chamblin}\ \emph {et~al.}(1999)\citenamefont
  {Chamblin}, \citenamefont {Emparan}, \citenamefont {Johnson},\ and\
  \citenamefont {Myers}}]{Chamblin:1999hg}%
  \BibitemOpen
  \bibfield  {author} {\bibinfo {author} {\bibfnamefont {A.}~\bibnamefont
  {Chamblin}}, \bibinfo {author} {\bibfnamefont {R.}~\bibnamefont {Emparan}},
  \bibinfo {author} {\bibfnamefont {C.~V.}\ \bibnamefont {Johnson}}, \ and\
  \bibinfo {author} {\bibfnamefont {R.~C.}\ \bibnamefont {Myers}},\ }\href
  {\doibase 10.1103/PhysRevD.60.104026} {\bibfield  {journal} {\bibinfo
  {journal} {Phys. Rev.}\ }\textbf {\bibinfo {volume} {D60}},\ \bibinfo {pages}
  {104026} (\bibinfo {year} {1999})}\BibitemShut {NoStop}%
\bibitem [{\citenamefont {Lee}\ \emph {et~al.}(2009)\citenamefont {Lee},
  \citenamefont {Park},\ and\ \citenamefont {Sin}}]{Lee:2009bya}%
  \BibitemOpen
  \bibfield  {author} {\bibinfo {author} {\bibfnamefont {B.-H.}\ \bibnamefont
  {Lee}}, \bibinfo {author} {\bibfnamefont {C.}~\bibnamefont {Park}}, \ and\
  \bibinfo {author} {\bibfnamefont {S.-J.}\ \bibnamefont {Sin}},\ }\href
  {\doibase 10.1088/1126-6708/2009/07/087} {\bibfield  {journal} {\bibinfo
  {journal} {JHEP}\ }\textbf {\bibinfo {volume} {07}},\ \bibinfo {pages} {087}
  (\bibinfo {year} {2009})}\BibitemShut {NoStop}%
\bibitem [{\citenamefont {Colangelo}\ \emph {et~al.}(2011)\citenamefont
  {Colangelo}, \citenamefont {Giannuzzi},\ and\ \citenamefont
  {Nicotri}}]{Colangelo:2010pe}%
  \BibitemOpen
  \bibfield  {author} {\bibinfo {author} {\bibfnamefont {P.}~\bibnamefont
  {Colangelo}}, \bibinfo {author} {\bibfnamefont {F.}~\bibnamefont
  {Giannuzzi}}, \ and\ \bibinfo {author} {\bibfnamefont {S.}~\bibnamefont
  {Nicotri}},\ }\href {\doibase 10.1103/PhysRevD.83.035015} {\bibfield
  {journal} {\bibinfo  {journal} {Phys. Rev.}\ }\textbf {\bibinfo {volume}
  {D83}},\ \bibinfo {pages} {035015} (\bibinfo {year} {2011})}\BibitemShut
  {NoStop}%
\bibitem [{\citenamefont {Andreev}\ and\ \citenamefont
  {Zakharov}(2007{\natexlab{b}})}]{Andreev:2006nw}%
  \BibitemOpen
  \bibfield  {author} {\bibinfo {author} {\bibfnamefont {O.}~\bibnamefont
  {Andreev}}\ and\ \bibinfo {author} {\bibfnamefont {V.~I.}\ \bibnamefont
  {Zakharov}},\ }\href {\doibase 10.1088/1126-6708/2007/04/100} {\bibfield
  {journal} {\bibinfo  {journal} {JHEP}\ }\textbf {\bibinfo {volume} {04}},\
  \bibinfo {pages} {100} (\bibinfo {year} {2007}{\natexlab{b}})}\BibitemShut
  {NoStop}%
\bibitem [{\citenamefont {Andreev}\ and\ \citenamefont
  {Zakharov}(2006)}]{Andreev:2006ct}%
  \BibitemOpen
  \bibfield  {author} {\bibinfo {author} {\bibfnamefont {O.}~\bibnamefont
  {Andreev}}\ and\ \bibinfo {author} {\bibfnamefont {V.~I.}\ \bibnamefont
  {Zakharov}},\ }\href {\doibase 10.1103/PhysRevD.74.025023} {\bibfield
  {journal} {\bibinfo  {journal} {Phys. Rev.}\ }\textbf {\bibinfo {volume}
  {D74}},\ \bibinfo {pages} {025023} (\bibinfo {year} {2006})}\BibitemShut
  {NoStop}%
\bibitem [{\citenamefont {Andreev}(2010)}]{Andreev:2010bv}%
  \BibitemOpen
  \bibfield  {author} {\bibinfo {author} {\bibfnamefont {O.}~\bibnamefont
  {Andreev}},\ }\href {\doibase 10.1103/PhysRevD.81.087901} {\bibfield
  {journal} {\bibinfo  {journal} {Phys. Rev.}\ }\textbf {\bibinfo {volume}
  {D81}},\ \bibinfo {pages} {087901} (\bibinfo {year} {2010})}\BibitemShut
  {NoStop}%
\bibitem [{\citenamefont {Bitaghsir~Fadafan}\ and\ \citenamefont
  {Tabatabaei}(2016{\natexlab{a}})}]{Fadafan:2015ynz}%
  \BibitemOpen
  \bibfield  {author} {\bibinfo {author} {\bibfnamefont {K.}~\bibnamefont
  {Bitaghsir~Fadafan}}\ and\ \bibinfo {author} {\bibfnamefont {S.~K.}\
  \bibnamefont {Tabatabaei}},\ }\href {\doibase 10.1103/PhysRevD.94.026007}
  {\bibfield  {journal} {\bibinfo  {journal} {Phys. Rev.}\ }\textbf {\bibinfo
  {volume} {D94}},\ \bibinfo {pages} {026007} (\bibinfo {year}
  {2016}{\natexlab{a}})}\BibitemShut {NoStop}%
\bibitem [{\citenamefont {Bitaghsir~Fadafan}\ and\ \citenamefont
  {Tabatabaei}(2016{\natexlab{b}})}]{Fadafan:2015kma}%
  \BibitemOpen
  \bibfield  {author} {\bibinfo {author} {\bibfnamefont {K.}~\bibnamefont
  {Bitaghsir~Fadafan}}\ and\ \bibinfo {author} {\bibfnamefont {S.~K.}\
  \bibnamefont {Tabatabaei}},\ }\href {\doibase 10.1088/0954-3899/43/9/095001}
  {\bibfield  {journal} {\bibinfo  {journal} {J. Phys.}\ }\textbf {\bibinfo
  {volume} {G43}},\ \bibinfo {pages} {095001} (\bibinfo {year}
  {2016}{\natexlab{b}})}\BibitemShut {NoStop}%
\bibitem [{\citenamefont {Buballa}(2005)}]{Buballa:2003qv}%
  \BibitemOpen
  \bibfield  {author} {\bibinfo {author} {\bibfnamefont {M.}~\bibnamefont
  {Buballa}},\ }\href {\doibase 10.1016/j.physrep.2004.11.004} {\bibfield
  {journal} {\bibinfo  {journal} {Phys. Rept.}\ }\textbf {\bibinfo {volume}
  {407}},\ \bibinfo {pages} {205} (\bibinfo {year} {2005})}\BibitemShut
  {NoStop}%
\bibitem [{\citenamefont {Schaefer}\ and\ \citenamefont
  {Wambach}(2005)}]{Schaefer:2004en}%
  \BibitemOpen
  \bibfield  {author} {\bibinfo {author} {\bibfnamefont {B.-J.}\ \bibnamefont
  {Schaefer}}\ and\ \bibinfo {author} {\bibfnamefont {J.}~\bibnamefont
  {Wambach}},\ }\href {\doibase 10.1016/j.nuclphysa.2005.04.012} {\bibfield
  {journal} {\bibinfo  {journal} {Nucl. Phys.}\ }\textbf {\bibinfo {volume}
  {A757}},\ \bibinfo {pages} {479} (\bibinfo {year} {2005})}\BibitemShut
  {NoStop}%
\bibitem [{\citenamefont {Cai}\ \emph {et~al.}(2012)\citenamefont {Cai},
  \citenamefont {He},\ and\ \citenamefont {Li}}]{Cai:2012xh}%
  \BibitemOpen
  \bibfield  {author} {\bibinfo {author} {\bibfnamefont {R.-G.}\ \bibnamefont
  {Cai}}, \bibinfo {author} {\bibfnamefont {S.}~\bibnamefont {He}}, \ and\
  \bibinfo {author} {\bibfnamefont {D.}~\bibnamefont {Li}},\ }\href {\doibase
  10.1007/JHEP03(2012)033} {\bibfield  {journal} {\bibinfo  {journal} {JHEP}\
  }\textbf {\bibinfo {volume} {03}},\ \bibinfo {pages} {033} (\bibinfo {year}
  {2012})}\BibitemShut {NoStop}%
\bibitem [{\citenamefont {Kaczmarek}\ \emph {et~al.}(2002)\citenamefont
  {Kaczmarek}, \citenamefont {Karsch}, \citenamefont {Petreczky},\ and\
  \citenamefont {Zantow}}]{Kaczmarek:2002mc}%
  \BibitemOpen
  \bibfield  {author} {\bibinfo {author} {\bibfnamefont {O.}~\bibnamefont
  {Kaczmarek}}, \bibinfo {author} {\bibfnamefont {F.}~\bibnamefont {Karsch}},
  \bibinfo {author} {\bibfnamefont {P.}~\bibnamefont {Petreczky}}, \ and\
  \bibinfo {author} {\bibfnamefont {F.}~\bibnamefont {Zantow}},\ }\href
  {\doibase 10.1016/S0370-2693(02)02415-2} {\bibfield  {journal} {\bibinfo
  {journal} {Phys. Lett.}\ }\textbf {\bibinfo {volume} {B543}},\ \bibinfo
  {pages} {41} (\bibinfo {year} {2002})}\BibitemShut {NoStop}%
\bibitem [{\citenamefont {Petreczky}\ and\ \citenamefont
  {Petrov}(2004)}]{Petreczky:2004pz}%
  \BibitemOpen
  \bibfield  {author} {\bibinfo {author} {\bibfnamefont {P.}~\bibnamefont
  {Petreczky}}\ and\ \bibinfo {author} {\bibfnamefont {K.}~\bibnamefont
  {Petrov}},\ }\href {\doibase 10.1103/PhysRevD.70.054503} {\bibfield
  {journal} {\bibinfo  {journal} {Phys. Rev.}\ }\textbf {\bibinfo {volume}
  {D70}},\ \bibinfo {pages} {054503} (\bibinfo {year} {2004})}\BibitemShut
  {NoStop}%
\bibitem [{\citenamefont {Kaczmarek}\ and\ \citenamefont
  {Zantow}(2006)}]{Kaczmarek:2005zp}%
  \BibitemOpen
  \bibfield  {author} {\bibinfo {author} {\bibfnamefont {O.}~\bibnamefont
  {Kaczmarek}}\ and\ \bibinfo {author} {\bibfnamefont {F.}~\bibnamefont
  {Zantow}},\ }\bibfield  {booktitle} {\emph {\bibinfo {booktitle}
  {{Proceedings, 23rd International Symposium on Lattice field theory (Lattice
  2005): Dublin, Ireland, Jul 25- 30,2005}}},\ }\href@noop {} {\bibfield
  {journal} {\bibinfo  {journal} {PoS}\ }\textbf {\bibinfo {volume}
  {LAT2005}},\ \bibinfo {pages} {192} (\bibinfo {year} {2006})},\ \Eprint
  {http://arxiv.org/abs/hep-lat/0510094} {arXiv:hep-lat/0510094 [hep-lat]}
  \BibitemShut {NoStop}%
\bibitem [{\citenamefont {Hashimoto}\ and\ \citenamefont
  {Kharzeev}(2014)}]{Hashimoto:2014fha}%
  \BibitemOpen
  \bibfield  {author} {\bibinfo {author} {\bibfnamefont {K.}~\bibnamefont
  {Hashimoto}}\ and\ \bibinfo {author} {\bibfnamefont {D.~E.}\ \bibnamefont
  {Kharzeev}},\ }\href {\doibase 10.1103/PhysRevD.90.125012} {\bibfield
  {journal} {\bibinfo  {journal} {Phys. Rev.}\ }\textbf {\bibinfo {volume}
  {D90}},\ \bibinfo {pages} {125012} (\bibinfo {year} {2014})}\BibitemShut
  {NoStop}%
\bibitem [{\citenamefont {Yang}\ and\ \citenamefont
  {Yuan}(2015)}]{Yang:2015aia}%
  \BibitemOpen
  \bibfield  {author} {\bibinfo {author} {\bibfnamefont {Y.}~\bibnamefont
  {Yang}}\ and\ \bibinfo {author} {\bibfnamefont {P.-H.}\ \bibnamefont
  {Yuan}},\ }\href {\doibase 10.1007/JHEP12(2015)161} {\bibfield  {journal}
  {\bibinfo  {journal} {JHEP}\ }\textbf {\bibinfo {volume} {12}},\ \bibinfo
  {pages} {161} (\bibinfo {year} {2015})}\BibitemShut {NoStop}%
\bibitem [{\citenamefont {Finazzo}\ and\ \citenamefont
  {Noronha}(2015)}]{Finazzo:2014rca}%
  \BibitemOpen
  \bibfield  {author} {\bibinfo {author} {\bibfnamefont {S.~I.}\ \bibnamefont
  {Finazzo}}\ and\ \bibinfo {author} {\bibfnamefont {J.}~\bibnamefont
  {Noronha}},\ }\href {\doibase 10.1007/JHEP01(2015)051} {\bibfield  {journal}
  {\bibinfo  {journal} {JHEP}\ }\textbf {\bibinfo {volume} {01}},\ \bibinfo
  {pages} {051} (\bibinfo {year} {2015})}\BibitemShut {NoStop}%
\bibitem [{\citenamefont {Liu}\ \emph {et~al.}(2007)\citenamefont {Liu},
  \citenamefont {Rajagopal},\ and\ \citenamefont {Wiedemann}}]{Liu:2006nn}%
  \BibitemOpen
  \bibfield  {author} {\bibinfo {author} {\bibfnamefont {H.}~\bibnamefont
  {Liu}}, \bibinfo {author} {\bibfnamefont {K.}~\bibnamefont {Rajagopal}}, \
  and\ \bibinfo {author} {\bibfnamefont {U.~A.}\ \bibnamefont {Wiedemann}},\
  }\href {\doibase 10.1103/PhysRevLett.98.182301} {\bibfield  {journal}
  {\bibinfo  {journal} {Phys. Rev. Lett.}\ }\textbf {\bibinfo {volume} {98}},\
  \bibinfo {pages} {182301} (\bibinfo {year} {2007})}\BibitemShut {NoStop}%
\bibitem [{\citenamefont {Avramis}\ \emph {et~al.}(2007)\citenamefont
  {Avramis}, \citenamefont {Sfetsos},\ and\ \citenamefont
  {Zoakos}}]{Avramis:2006em}%
  \BibitemOpen
  \bibfield  {author} {\bibinfo {author} {\bibfnamefont {S.~D.}\ \bibnamefont
  {Avramis}}, \bibinfo {author} {\bibfnamefont {K.}~\bibnamefont {Sfetsos}}, \
  and\ \bibinfo {author} {\bibfnamefont {D.}~\bibnamefont {Zoakos}},\ }\href
  {\doibase 10.1103/PhysRevD.75.025009} {\bibfield  {journal} {\bibinfo
  {journal} {Phys. Rev.}\ }\textbf {\bibinfo {volume} {D75}},\ \bibinfo {pages}
  {025009} (\bibinfo {year} {2007})}\BibitemShut {NoStop}%
\end{thebibliography}%

\end{document}